\documentclass[review]{elsarticle}
\usepackage{lineno,hyperref,eurosym}
\usepackage{color}
\usepackage{tabularx}
\usepackage[utf8]{inputenc}
\usepackage{graphicx}
\usepackage{amsmath}
\usepackage{lscape}
\usepackage{xcolor, soul}
\usepackage{makecell}

\journal{Nonlinear dynamics}
\begin{document}
\begin{frontmatter}
\title{Uncontrolled geostationary satellites: mapping periodic transitions to chaos with Lagrangian Descriptors}
\author[label1]{R. Flores}
\author[label2]{J. Daquin}
\author[label3]{M. Pontani}
\author[label4]{H. Susanto}
\author[label1,label5,label6]{E. Fantino$^{\star}$}
\address[label1]{Department of Aerospace Engineering, Khalifa University of Science and Technology, P.O. Box 127788, Abu Dhabi (United Arab Emirates)}
\address[label2]{Universit\'e de Toulon, Aix Marseille Univ, CNRS, CPT, Toulon (France)}
\address[label3]{Department of Astronautical, Electrical, and Energy Engineering, Sapienza University of Rome, via Salaria 851, 00138 Rome (Italy)}
\address[label4]{Department of Mathematics, Khalifa University of Science and Technology, P.O. Box 127788, Abu Dhabi (United Arab Emirates)}
\cortext[cor]{Corresponding author: elena.fantino@ku.ac.ae (E.~Fantino)}
\address[label5]{Polar Research Center, Khalifa University of Science and Technology, P.O. Box 127788, Abu Dhabi (United Arab Emirates)}
\address[label6]{Khalifa University Space Technology and Innovation Lab, Khalifa University of Science and Technology, P.O. Box 127788, Abu Dhabi (United Arab Emirates)}

\begin{abstract}
Uncontrolled geostationary satellites abandoned near an unstable equilibrium point of the equator experience irregular transitions between dynamical states (continuous circulation, long and short libration). They are caused by the interaction between the longitudinal dynamics, governed by the tesseral harmonics of the geopotential, and the orbital precession forced by Earth's oblateness and lunisolar perturbations. The transitions are extremely sensitive to small perturbations, making the long-term evolution unpredictable. Recently, a Monte Carlo analysis of trajectories starting in the immediate vicinity of the $165^\circ$E unstable equilibrium point, revealed that the evolution to chaos is not gradual. It occurs via sudden episodes of disorder at specific points of the precession cycle, when the orbital inclination is minimal. Due to the high cost of the statistical analysis, the results  where limited to a single initial longitude. This paper applies modified versions of the diameter Lagrangian descriptor to reduce the computational burden. This enables mapping the dynamical behavior over the complete range of longitudes where transitions between modes of motion are possible, considering both unstable equilibrium points ($165^\circ$E and $15^\circ$W). It is found that the episodes of chaos remain linked to the orbital inclination cycle, but their timing depends on the initial spacecraft longitude. As the initial position moves farther away from the unstable point, the transitions take place at higher values of the orbital inclination. The longitudes where the transitions occur at maximum inclination correspond to the boundaries of the chaotic region.
\end{abstract}


\end{frontmatter}


\section{Introduction}
The proliferation of space debris is a source of major concern for the sustainability of space operations. The geostationary orbit (GEO) is of special concern, due to its outstanding operational interest and its level of congestion. Nowadays, disposal to a graveyard orbit is the standard approach for GEO spacecraft at the end of operations. However, the disposal maneuver is not always successful. Moreover, old satellites were often abandoned at their operational station. Given the potential risk for the operations of current and future spacecraft, the long-term dynamics of decommissioned GEO spacecraft is an active area of research \cite{Colombo_2017}\cite{Celletti_2014}.

Two notable features of the evolution of resident space objects (RSO) in geostationary orbit are the precession of the orbital plane and secular changes of the geographical longitude ($\lambda$).
The precession is driven by the coupling of Earth’s oblateness with lunisolar perturbations \cite{Allan_1964}\cite{Ha_1986}. Hechler \cite{Hechler_1985} reported a period of 53 years for the inclination cycle, with the precession axis inclined $7.3^\circ$ from the Earth's poles, corresponding to an inclination amplitude of $14.6^\circ$. In long-term numerical propagations, Friesen et al. \cite{Friesen_1992} observed inclination variations reaching 15º over a 53-year cycle.

The drift in longitude is caused by the non-uniformity of Earth’s gravity field. The $J_{22}$ sectorial harmonic, which quantifies the ellipticity of the equator, plays a dominant role due to the resonance between the periods of the orbit and Earth’s rotation. It creates two stable ($75^\circ$E and $108^\circ$W) and two unstable ($165^\circ$E and $15^\circ$W) equilibrium points along the equator \cite{Blitzer_1962}\cite{Blitzer_1965}\cite{Musen_1962}. Note that these longitudes are only approximate. Different values can be found in the literature, depending on the physical model used by each author (i.e., the number of terms included in the harmonic synthesis of the geopotential).

The resulting motion can be short libration (oscillation around a stable longitude, SL hereafter) or continuous circulation (monotonic east or west drift, CC). Higher-degree harmonics of the geopotential introduce an asymmetry between the unstable equilibrium points that enables an additional motion type, long libration (LL) spanning both stable points \cite{Vashkoy_1973}\cite{Kiladze_1997a}\cite{Kiladze_1997b}.

The interaction between the precession of the orbital plane and the tesseral harmonics governing longitudinal motion modulates the potential barrier at the unstable equilibrium points. As a result, the trajectory may be able to traverse them sporadically, transitioning between different modes of motion \cite{Kiladze_1997a}. Breiter et al. \cite{Breiter_2005} studied the stability of geostationary and super-geostationary (higher altitude graveyard orbits) RSO over a period of 40 years. A small fraction of the geosynchoronous orbits that they analyzed showed signs of chaotic behavior. They where associated with the separatrix between the CC and LL regions around $165^\circ$E.

Proietti et al. \cite{Proietti_2021} performed a high-fidelity analysis of the long-term evolution of GEO spacecraft abandoned near the unstable points. It revealed that the occurrence of CC/LL and LL/SL transitions is very sensitive to minute perturbations (both physical and numerical). So much that, for some initial longitudes the trajectory is effectively unpredictable beyond a horizon of approximately 50 years. To improve the characterization of this irregular behavior, \cite{Flores_2023} presented a Monte Carlo simulation of clouds of spacecraft released very close to the $165^\circ$E unstable point, propagating their trajectories for 120 years. The study unveiled sudden increases in the scatter of the clouds, interspersed with intervals of smooth evolution. The short episodes of chaos, triggered by transitions between LL and CC, have a periodicity of approximately half a century, and are linked to the precession of the orbital plane.

While the Monte Carlo analysis from \cite{Flores_2023} succeeded in unveiling the cyclic disorder episodes, it has important shortcomings. It can only explore one point along the equator at a time, and requires a substantial post-processing effort. For this reason, only the $165^\circ$E point was analyzed. The study did not explore if the same phenomenon occurs for other initial positions. Furthermore, the technique is not well suited to investigate the evolution beyond one century, because the bundle of trajectories disperses all over the equator and it becomes difficult to measure subsequent changes in scatter. Thus, the evolution over very long time scales has not been characterized. Finally, due to the requirement to propagate a large number of trajectories for each initial position, the method is computationally onerous and demands substantial storage capacity. Thus, the approach would be impractical to investigate the dynamical behavior over wide ranges of initial longitudes.

This paper presents the first comprehensive characterization of the chaotic transitions between modes of motion. It maps their timing as a function of the initial position, covering the complete range of longitudes where irregular behavior is possible (i.e., the neighborhood of both unstable equilibrium points). To address the limitations of the Monte Carlo technique, the system dynamics is characterized using Lagrangian descriptors (LD). This method was originally developed to study the phase space structure of geophysical flows \cite{Madrid_2009}\cite{Mendoza_2010}. The strategy also proved effective in chemistry \cite{Craven_2015}\cite{Craven_2016}\cite{Craven_2017}\cite{Revuelta_2019}. In recent times, it has found applications in the domain of astrodynamics \cite{Gkolias_2016}\cite{Daquin_2023}.

The LD method starts with a grid of initial conditions over a section of the phase space. The corresponding trajectories are propagated for a fixed time, integrating a norm of the trajectory (e.g., arc length \cite{Mancho_2013}) along the path. Variations of the integral correlate to dynamical structures of the phase space. The method is very general, it can be applied to autonomous as well as time-dependent systems and does not rely on the existence of a Hamiltonian. This strategy is markedly different from alternatives like Lyapunov exponents \cite{Villac_2008}\cite{Balcerzak_2018}\cite{Canales_2023}\cite{Fiedler_2024}, because LD always accumulate a positive scalar quantity, of geometric or dynamic character, along the trajectories.

The standard approach in the literature is to evaluate the LD over a fixed time interval. However, in this study, it is crucial to extract information about the timing of the transitions. To this end, the LD are computed over a moving time window. Its span is tailored to the period of the librational motions, in order to discriminate reliably between the different types dynamical response while preserving sufficient resolution in time.
Leveraging LD, it is possible to explore the behavior of objects abandoned anywhere along the GEO ring, with reduced computational cost and minimal post-processing effort. In stark contrast to the Monte Carlo analysis, LD only require one orbit propagation per initial longitude. Furthermore, the dynamics can be easily explored over time scales of several centuries.

This paper is structured as follows: Sec.~\ref{sec_dyn_geo} provides additional details about the dynamics of GEO satellites near an unstable point; Sec.~\ref{sec_mat_meth} introduces the numerical model and the application of LD to the problem at hand; Sec.~\ref{sec_results} presents the numerical results for both unstable equilibrium points; finally, the main conclusions are drawn in Sec.~\ref{sec_concl}.

\section{The dynamics of geostationary RSO near an unstable equilibrium point}
\label{sec_dyn_geo}

\begin{figure}[h!]
	\centering
	\includegraphics[width=0.7\textwidth]{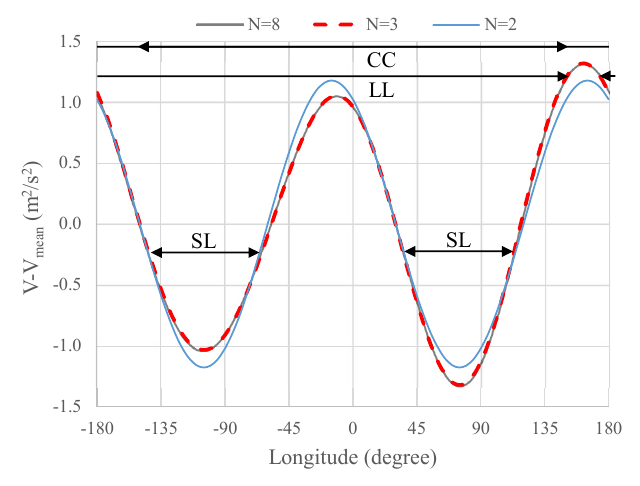}
	\caption{Potential along equator vs. geopotential expansion degree (N).}
	\label{fig:eq_pot}
\end{figure}

In a first approximation, the longitudinal dynamics can be described qualitatively by considering only harmonics of the gravitational field up to degree 2. The $J_{22}$ term, the degree-2 sectorial harmonic representing the ellipticity of the equator, gives rise to four equilibrium points (blue line in Fig.~\ref{fig:eq_pot}). Two are stable ($S_1$ and $S_2$) and two unstable ($U_1$ and $U_2$) \cite{Blitzer_1962}\cite{Blitzer_1965}\cite{Musen_1962}. The figure uses the sign convention that the gravity acceleration is the gradient of the potential (V). Thus, the longitudinal component of the gravitational perturbation is attractive at $U_1$ and $U_2$ (the unstable points are on the major axis of the equator, where the mass is concentrated) and repulsive at the stable points. This counter-intuitive fact is due to the changes in energy that the spacecraft experiences, which alter the orbital period and affect the motion relative to Earth. For example, a spacecraft approaching $U_1$ from the East experiences a westward pull. This reduces its orbital velocity in an Earth-centered non-rotating frame, causing it to drop to a lower orbit. The orbital period becomes shorter, and the satellite overtakes Earth's spin. Therefore, in a planet-fixed frame, it accelerates towards the East (away from $U_1$).

Under the influence of $J_{22}$ alone, the motion of a RSO would be a short libration encompassing only the closest stable point, or continuous circulation, depending on the initial conditions. The third-degree sectorial and tesseral harmonics of the geopotential break the symmetry of the unstable points (red dashed curve in Fig.~\ref{fig:eq_pot}) \cite{Kiladze_1997a}. They give rise to an additional mode of motion: long libration encompassing both stable points. Higher-degree harmonics of the gravity field have a minor effect on the potential distribution along the equator (see the black line in Fig.~\ref{fig:eq_pot}, which corresponds to a model complete to degree and order 8).

Additionally, Earth’s oblateness (measured by the $J_2$ zonal harmonic) forces the orbital plane to precess around the celestial pole. At the same time, third-body perturbations (lunar and solar gravity) cause a precession around the pole of the orbit of the perturbing object. The combined action of $J_2$ and lunisolar perturbations yields a 53-year inclination cycle. The axis of precession is the pole of the Laplacian plane, intermediate between Earth's axis and the pole of the ecliptic. It forms an approximate angle of $7^\circ$ with the celestial pole \cite{Allan_1964}\cite{Grigoriev_1993}, but this value changes continuously as the lunar orbit precesses.

The net result is that the orbital inclination of GEO RSO oscillates between $0^\circ$ and $14^\circ$ with a period of 53 years. The coupling between this inclination cycle and the longitudinal dynamics affects the potential barrier required to move across the unstable points (it could be visualized as a vertical oscillation of the black curve in Fig.~\ref{fig:eq_pot}). The barrier is highest at the point of zero inclination of the precession cycle. The unstable points act as gates that open and close sporadically, enabling RSO to transition between different modes of motion \cite{Gazzino_2017}. For $U_2$, the transition is between CC and LL, while for $U_1$ it involves changes between LL and SL. The amplitude of the potential barrier fluctuation is small, preventing transitions between CC and SL modes. 

The reader must keep in mind that there are no true equilibrium points along the equator. For GEO, the magnitude of lunisolar perturbations and the irregularities of the geopotential are comparable \cite{Friesen_1992}. Given that the former depend on the instantaneous configuration of the Sun-Earth-Moon system, any equilibrium position can only be temporary. For the sake of convenience, it is still useful to talk about equilibrium points. However, what they really denote is a small region surrounding the theoretical longitude where the centrifugal and gravitational forces cancel each other in an idealized model.

A high-fidelity study of a small sample of trajectories starting near both unstable points \cite{Proietti_2021} demonstrated that there are initial positions along the equator for which the trajectory propagation does not converge beyond 60 years. That is, as the integrator tolerance is ramped down, the long-term solution fluctuates wildly instead of converging to an asymptotic value. This occurs because the system is overly sensitive. The discretization and rounding errors of the numerical integrator are enough to change the long-term outcome. Repeating the calculations with quadruple precision arithmetic and very small integration tolerances did not solve the issue, evidencing the unpredictability of the behavior in practice. 

To further characterize the evolution towards unpredictability, \cite{Flores_2023} used a Monte Carlo analysis of tightly packed clouds of hundreds of spacecraft released close to $U_2$ and propagated for 120 years. The uncertainty of the trajectories was quantified monitoring the scatter of the clouds. A GEO satellite abandoned near $U_2$ has almost exactly the energy required to overcome the potential barrier during the initial epoch (when the inclination is null). Therefore, over most of the precession cycle, when the barrier is lower, it moves in CC. Once every 53 years, at the points of null inclination of the precession cycle, it comes to a near-complete stop (in an Earth-fixed frame) at $U_2$. How long the satellite dwells at the equilibrium point, and whether or not it transitions between LL and CC, are extremely sensitive to te state of the spacecraft and the timing of arrival to $U_2$ (which affects third-body perturbations). This causes the scatter of the clouds to increase by an order of magnitude in a very short time ($\sim$2 years). Once the spacecraft leave the vicinity of $U_2$, the motion becomes regular and the scatter stabilizes (subsequent approaches to $U_2$ do not increase the dispersion). After two episodes of disorder (i.e., approximately half a century) the scatter is so large that the clouds spread all around the equator. This indicates that the trajectories become effectively unpredictable. Awareness of the timing of the transitions is crucial to schedule updates of the orbit determination for GEO debris, because they represent horizons beyond which orbit propagation becomes unreliable.

These periodic episodes of disorder were discovered only recently, since they are not easy to identify in the phase space structure of the problem. They can occur only at very specific points in time, when the spacecraft comes to a stop at the instantaneous equilibrium point. Here, "instantaneous" indicates that, due to lunisolar perturbations, there are no true fixed equilibrium points along the equator, because the force balance at a given longitude changes with time (i.e., the system is not autonomous). Moreover, the motion during consecutive passages through the same point is different due to changes in the relative position of the Sun, Earth and Moon. The traditional approach with averaging techniques fails to capture these nuances, making the phenomenon difficult to detect.

Another important finding of \cite{Flores_2023} is that the main features of this phenomenon can be reproduced retaining only gravity harmonics up to degree 3 and lunisolar perturbations with a simple analytical model (Earth and Moon in circular coplanar orbits). Using a higher fidelity model added nuances to the behavior (e.g., variations of the magnitude of the jumps in scatter), but did not change the overall qualitative response. 

\section{Material and methods}
\label{sec_mat_meth}

\subsection{Numerical modeling}
\label{sec_num_mod}
The trajectory is propagated in Cartesian coordinates in the non-rotating Geocentric Celestial Reference System (GCRS) \cite{Hilton_2012}. The equations of motion (summarized in Appendix A, for reference) are integrated directly, without any averaging. Numerical integration relies on an explicit adaptive embedded Runge-Kutta scheme of orders 9(8) with order 8 dense output \cite{Verner_RK}. The forces considered are Earth, Moon and Sun gravity, and solar radiation pressure.

Gravity calculations follow the guidelines from the International Earth rotation and Reference Systems Service (IERS) Technical Note No. 36 \cite{IERS_2010a}. It designates EGM2008 \cite{Pavlis_2012} as the geopotential model of reference, provides an improved value of $J_2$ and adds secular variations of the low-degree zonal harmonics. The zero-tide version of the gravity model has been used. The harmonic synthesis of the gravitational acceleration is computed with the modified forward row recursion scheme \cite{Holmes_2002}. The maximum degree of harmonic synthesis in the calculations is $N=8$ (i.e., the series is complete to degree and order 8). According to \cite{Flores_2021}, $N=8$ delivers the maximum accuracy achievable by EGM2008 at GEO altitude. Adding harmonics of higher degree does not improve the results, because the uncertainty of the model coefficients becomes dominant.

The spacecraft state is formulated in GCRS, but the geopotential is expressed in the Earth-fixed Terrestrial Reference System (TRS). The transformation between the two systems is implemented with the IAU 2000/2006 combined precession-nutation model \cite{Hilton_2006}. Instead of the complete model, a simplified formulation based on Ref. \cite{Capitaine_2008} is used. It maintains an accuracy of 1 arcsec over periods of 500 years, while reducing the computational cost substantially. Polar motion is not included in the calculations, because it cannot be predicted accurately \cite{Hilton_2012}. Its effect is on the order of 0.3 arcsec, below the uncertainty of the simplified precession-nutation model. The Earth Rotation Angle (ERA) is computed according to \cite{IERS_2010b}.

The calculation of the lunisolar perturbations uses precomputed ephemeris from the JPL Horizons system \cite{JPL_hor}, interpolated with cubic splines. The effect of rounding errors is minimized with the well-conditioned expression of the tidal forces from Ref.~\cite{Battin_1999}.

The acceleration due to solar radiation pressure is computed assuming a spherical spacecraft with 100\% specular reflectivity. An area-to-mass ratio of $7 \times 10^{-3}$ m$^2$kg$^{-1}$ has been selected, typical of a communications satellite. The eclipses are modeled with a double-cone model (umbra+penumbra), as described in \cite{Flores_2021}. The dependence of solar flux with Sun-spacecraft distance is included in the calculation.

Flores et al. \cite{Flores_2023} demonstrated that the main features of the motion of GEO satellites released in the immediate vicinity of $U_2$ can be reproduced with a simplified physical model. It only considers gravity harmonics up to the third degree ($N=3$) and lunisolar perturbations approximated with circular coplanar orbits for Earth and Moon (i.e., the Moon lies on the plane of the ecliptic). This approximation neglects changes of orientation of Earth's axis, as well as radiation pressure. In spite of the vastly simplified dynamics, the transitions between the two modes of motion remain unpredictable, and occur at similar epochs. A comparison of the results from the two levels of fidelity will be presented, to demonstrate that they yield similar behavior over a wide range of initial longitudes.

\subsection{Mapping the changes in motion type using Lagrangian descriptors}
Let the state vector of the RSO be ${\bf x} = [{\bf r}, {\bf v}]^T$, where $\bf r$ denotes position and $\bf v$ velocity. The ordinary differential equation governing the trajectory is (see the Appendix for more details)
\begin{equation} \label{eq_motion}
	{\bf \dot x} = {\bf f} ({\bf x}, t) \quad , \quad t \in [t_i, t_f],
\end{equation}
where ${\bf f} = [{\bf v}, {\bf a}]^T$ and $\bf a$ is the acceleration in GCRS. Equation \ref{eq_motion} is supplemented with the initial condition
\begin{equation} \label{eq_inicon}
	{\bf x} (t_i) = {\bf x}_i.
\end{equation}
Integrating the differential equation together with the initial condition gives the solution ${\bf x} ({\bf x}_i , t)$.

In its original form \cite{Mendoza_2010}, the LD is simply the Euclidean length of the trajectory. The rationale is that trajectories starting from neighboring points but displaying markedly different dynamical behavior (e.g., particles moving on opposite sides of a shear layer) yield vastly different values of the arc length. Thus, sharp gradients of the LD correspond to transitions in the dynamical structure of the system. For the same reason, instead of the speed, one could integrate the magnitude of the acceleration or the curvature to detect changes in the dynamical structure of the system \cite{Mancho_2013}. In general, any scalar function of the state $\psi({\bf x})$ (designated as observable henceforth) can be used to define a LD.

Another related and extremely simple approach is to characterize the trajectory based on its extent in phase space. The idea was introduced by Dvorak as the Maximum Eccentricity Method \cite{Dvorak_2004}, to analyze the stability of exoplanet orbits. The concept has been applied successfully in other fields \cite{Mundel_2014}, and is known as the diameter (or amplitude) Lagrangian diagnostic. Given an observable $\psi$, the amplitude of a trajectory is computed as
\begin{equation} \label{eq_LD_dia_basic}
	\mathcal{D}_\psi ({{\bf x}_i}) = {\max}_\tau (\psi ({\bf x}_i, \tau)) - {\min}_\tau (\psi ({\bf x}_i, \tau ) ) \quad , \quad \tau \in [t_i, t_f].
\end{equation}
Trajectories with distinct dynamical response are expected to yield different values of the amplitude. While this indicator is simple to implement and inexpensive to evaluate, it is a function of the initial condition only. It does not contain any information about possible changes in the dynamics between the initial and final states. This is a serious limitation for the problem under study, because the objective is to identify, for each initial condition, the epochs when chaotic behavior can develop.  

We propose an extension of the diameter LD that retains information in the time domain. Given a time window $\left[ {t,t + \delta } \right] \subset \left[ {{t_i},{t_f}} \right]$, where $\delta > 0$ is the span, the diameter of a trajectory over the window span is defined as
\begin{equation} \label{eq_LD_dia}
	\Delta_\psi ({{\bf x}_i, t, \delta}) = {\max}_\tau (\psi ({\bf x}_i, \tau)) - {\min}_\tau (\psi ({\bf x}_i, \tau ) ) \quad , \quad \tau \in [t, t+\delta].
\end{equation}
By virtue of the moving time window, this generalization of the amplitude LD is able to characterize the motion at different epochs. Therefore it can determine when the dynamical state changes. Note that the traditional diameter \ref{eq_LD_dia}, when plotted in 2D, yields a dynamical map where both axes represent coordinates of the initial condition. In the modified indicator \ref{eq_LD_dia}, one axis is time ($\delta$ is kept fixed). 

For geostationary orbits, a specially relevant observable is the semimajor axis of the osculating ellipse ($a$). It has a direct relationship with the direction of longitudinal motion in an Earth-fixed frame. When $a$ is above the geosynchronous orbit value (i.e., higher orbit) the mean motion is slower than Earth's rotation. The planet overtakes the orbiter and it drifts westward in the rotating frame. Conversely, low values of $a$ correspond to eastward drift. Thus, librational motion alternates between high and low values of $a$, and is characterized by a larger diameter than continuous circulation.

\begin{figure}
	\centering
	\includegraphics[width=0.80\textwidth]{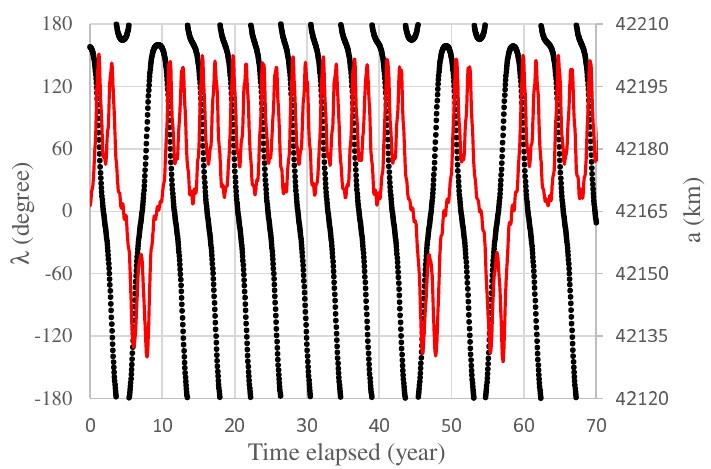}
	\caption{Longitude (black dots) \& semimajor axis (red line) vs. time. $\lambda_i=158^\circ$, $t_i=$1/1/2020, N=3.}
	\label{fig_avslon}
\end{figure}

Figure~\ref{fig_avslon} displays the evolution of $a$ and $\lambda$ (geographical longitude) for an object abandoned near $U_2$. The inclination of an uncontrolled GEO\footnote{Strictly speaking, the orbit remains approximately geostationary only if station-keeping maneuvers are applied regularly. Once abandoned, the RSO will move continuously in an Earth-fixed frame. For the sake of convenience, we will continue referring to it as geostationary, even in the uncontrolled state.} changes continuously due to precession. As a result of the nonzero inclination, the angular velocity of the orbiter along the equator varies across the orbit. It is minimal when the spacecraft crosses the equator, and maximal at the points of extreme latitude. In an Earth-fixed frame, the diurnal fluctuations of latitude and longitude create a figure eight pattern (analemma, see \cite{Chalmers_1987}). To remove this distracting effect from the plot, Fig.~\ref{fig_avslon} uses stroboscopic sampling. The longitude is saved at integer multiples of Earth's rotation period (one stellar day), hiding the diurnal modulation.

The ideal orbital radius of a GEO, assuming a spherical mass distribution for Earth, is 
\begin{equation} \label{eq_r_GEO}
	r = \left(\frac{\mu_E}{\Omega_E^2}\right) ^ {1/3},
\end{equation}
where $\mu_E$ is Earth's gravitational parameter (398600.4415 km$^3$s$^{-2}$ \cite{Pavlis_2012}) and $\Omega_E$ denotes its angular velocity in inertial space (1.00273781191135448 revolutions each 86400 seconds \cite{IERS_2010b}). One must keep in mind that, for a RSO rotating in sync with Earth, the altitude where the radial component of gravity cancels the centrifugal force is different from the value given by Eq.~\ref{eq_r_GEO}. Using the ideal radius (which does not account for the irregularities of the gravity field) yields a a slightly different orbital period. It is desirable to have initial conditions as close to geosynchronous as possible, to avoid obfuscating the behavior under study. Therefore, the spacecraft is placed over the equator, at rest in the rotating frame, with the initial height determined by a nonlinear solver that balances the radial component of the nonuniform gravity field with the centrifugal acceleration. This yields a slightly different initial radius for each longitude. For example, in the neighborhood of $U_2$ the corrected height is, on average, 600 m above the value from Eq.~\ref{eq_r_GEO}.

\begin{figure}
	\centering
	\includegraphics[width=0.80\textwidth]{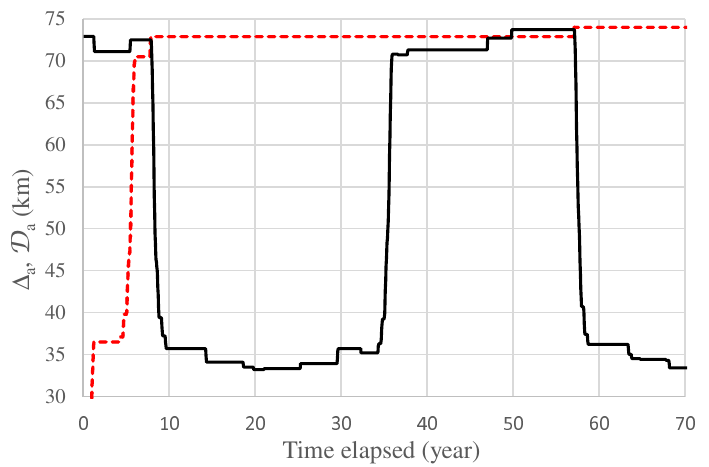}
	\caption{$\Delta_a$ with 10-year window (continuous black line) and $\mathcal{D}_a$ accumulated since start of propagation (dashed red) vs. time. $\lambda_i=158^\circ$, $t_i=$1/1/2020, N=3.}
	\label{fig_da_10yr_tot}
\end{figure}

Figure~\ref{fig_avslon} shows that, during continuous circulation, the semimajor axis fluctuates less than 40 km. On the other hand, during episodes of long libration (circa 5 and 50 years in the figure), the range of variation is close to 70 km. Hence, the diameter of $a$ is a good indicator of the mode of motion. However, the standard way of computing the diameter (i.e., over the complete duration of the propagation), cannot detect changes in dynamical state after the first episode of LL. This is illustrated in Fig.~\ref{fig_da_10yr_tot}, which compares the evolution of the diameter computed with a 10-year window ($\Delta_a$, solid black line) against the accumulated value ($\mathcal{D}_a$, dashed red), for the same trajectory depicted in Fig.~\ref{fig_avslon}. The windowed approach is able to detect multiple transitions between LL and CC. Furthermore, the distribution of $\Delta_a$ is strongly bimodal. This makes it easy to determine the mode of motion from the value of the amplitude. On the other hand, the accumulated diameter increases monotonically until it reaches the value corresponding to LL, and then remains constant. Therefore, the only information it provides is that, at the beginning of the propagation, there has been a period of librational motion.

\begin{figure}
	\centering
	\includegraphics[width=0.80\textwidth]{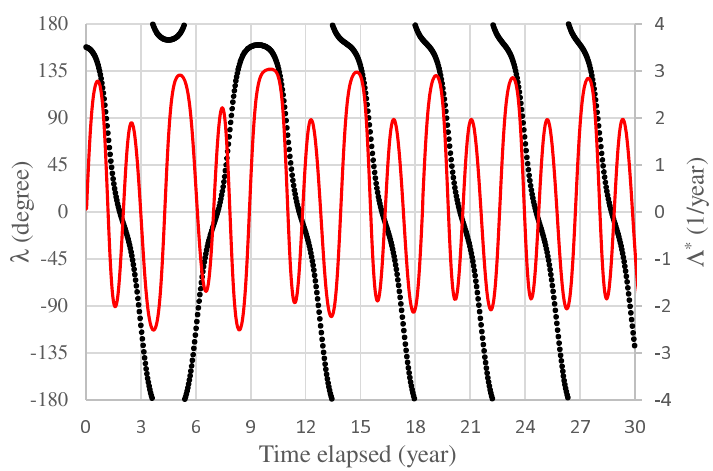}
	\caption{Largest Lyapunov exponent averaged over 15 orbits (red line) and longitude (black dots) vs. time. $\lambda_i=158^\circ$, $t_i=$1/1/2020, N=3.}
	\label{fig_lon_maxlya}
\end{figure}

It is interesting to compare the performance of proposed LD against the widely used largest Lyapunov exponent chaos indicator (LLE, denoted $\Lambda^*$ to avoid confusion with the geographic longitude). The LLE has been computed with the methodology presented in \cite{Dabrowski_2012}. Figure~\ref{fig_lon_maxlya} contains a detail of the results for the first 30 years of the trajectory shown in Fig.~\ref{fig_avslon}. The graph shows the time average of $\Lambda^*$ over 15 orbital periods (i.e., the simple moving average over 15 stellar days, approximately $359^\text{h} \, 1^\text{m} \, 1.48^\text{s}$). For the sake of convenience, the LLE calculation uses an Earth-fixed frame, so the spacecraft displacement over 15 orbits is actually small ($\sim1^\circ$ or less). There is a strong correlation between $\Lambda^*$ and the longitude. The positive and negative peaks of the LLE correspond to the motion between the unstable ($165^\circ$E and $15^\circ$W) and stable ($75^\circ$E and $108^\circ$W) equilibrium points.
When two neighboring trajectories approach an unstable point, the first one to arrive starts slowing down, allowing the other to catch up. Thus, their mutual distance decreases, corresponding to a negative Lyapunov exponent. Conversely, the first trajectory to depart from the unstable point accelerates towards one of the adjacent stable positions, and leaves the other behind. The separation increases and $\Lambda^*$ becomes negative. Reference \cite{Flores_2023} discusses this behavior in detail. The pattern repeats continuously, with a period of $\sim$2.5 years between positive peaks of $\Lambda^*$.

Note that, most of the time, the positive LLE peaks are not associated with chaotic behavior, because the changes in separation are reversible. For the example shown in Fig.~\ref{fig_lon_maxlya}, chaos only occurs circa 10 years after the initial epoch (when there is a transition between LL and CC), and only lasts for a short time ($\sim$2 years). The subsequent motion is regular, albeit with a larger separation between neighboring trajectories. The scatter grows irreversibly during the chaos episode, but it remains stable afterwards, until another transition occurs. See \cite{Flores_2023} for a comprehensive explanation.

\begin{figure}
	\centering
	\includegraphics[width=0.80\textwidth]{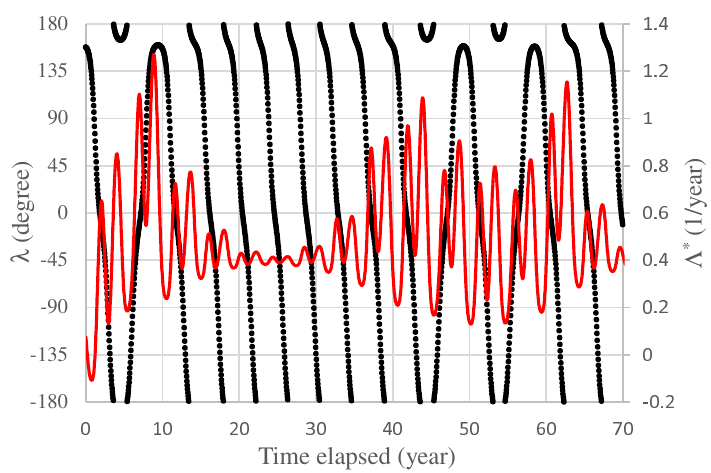}
	\caption{Largest Lyapunov exponent averaged over 1500 orbits (red line) and longitude (black dots) vs. time. $\lambda_i=158^\circ$, $t_i=$1/1/2020, N=3.}
	\label{fig_lon_maxlya_1500}
\end{figure}

To reveal the timing of the chaotic events, it is necessary to identify the transitions between LL and CC. To this end,
Fig.~\ref{fig_lon_maxlya_1500} uses a LLE averaging period of 1500 orbits\footnote{1500 orbits seems to be the sweet spot. Shorter and longer averaging periods make the changes in dynamical state less conspicuous.} ($\sim$4 years). With this setup, the height of the $\Lambda^*$ peaks increases near the epochs when the mode of motion changes. However, the signal is much noisier than in Fig.~\ref{fig_da_10yr_tot}. Thus, $\Delta_a$ proves superior in this respect, while being half as expensive to evaluate (LLE requires two propagations for each initial longitude, one for the trajectory and another for the perturbation).

The long-term (100 years) average of $\Lambda^*$ is 0.5 year$^{-1}$. The Lyapunov time $t_L=1/\Lambda^*$ is often considered a rough estimate of the limit of trajectory predictability. However, for this problem, its value (2 years) is much shorter than the actual threshold of unpredictability ($\sim$60 years \cite{Proietti_2021}). Interestingly, $t_L$ is comparable to the duration of the jumps in scatter observed in \cite{Flores_2023}.

\section{Results and discussion}
\label{sec_results}

\subsection{The dynamical behavior around $U_2 \: (165^\circ \rm{E})$}
In order to detect LL motion reliably, the window span used to evaluate $\Delta_a$ must be sufficient to capture the minimum and maximum extremes of the semimajor axis. Thus, the window should be longer that one half of the LL period  which, from Fig.~\ref{fig_avslon}, is close to 10 years (the precise value is a function of the initial longitude). On the other hand, to identify with precision the timing of the transitions between LL and CC, the window should be as narrow as possible. Thus, as a first guess, the window width should be between 5 and 10 years.

To find the optimal window span, a collection of GEO orbits starting in the vicinity of $U_2$ was propagated for 120 years, enough to capture two complete precession cycles. The initial longitudes range from 153$^\circ$E to 171$^\circ$E, covering the interval where transition between CC and LL are possible.

Because lunisolar perturbations change with time, the starting epoch affects the outcome of the simulations. For the initial tests, the propagation begins on 1 January 2020 at 0:00 UTC (JDN 2458849.5). This is one of the dates analyzed in \cite{Proietti_2021} and \cite{Flores_2023}, allowing direct comparison with the previous studies. 

\begin{figure}[h!]
	\centering
	\includegraphics[width=0.49\textwidth]{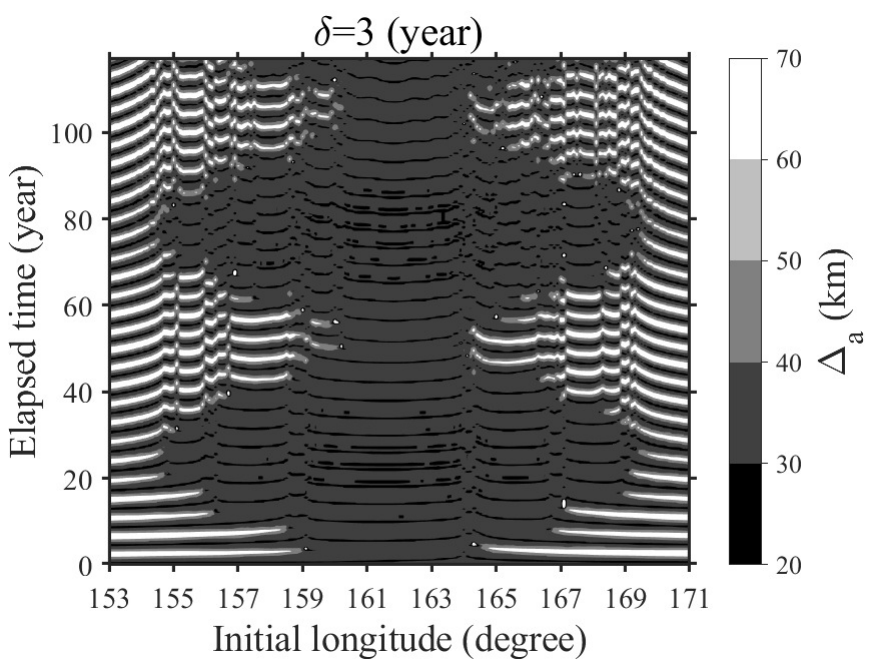}
	\includegraphics[width=0.49\textwidth]{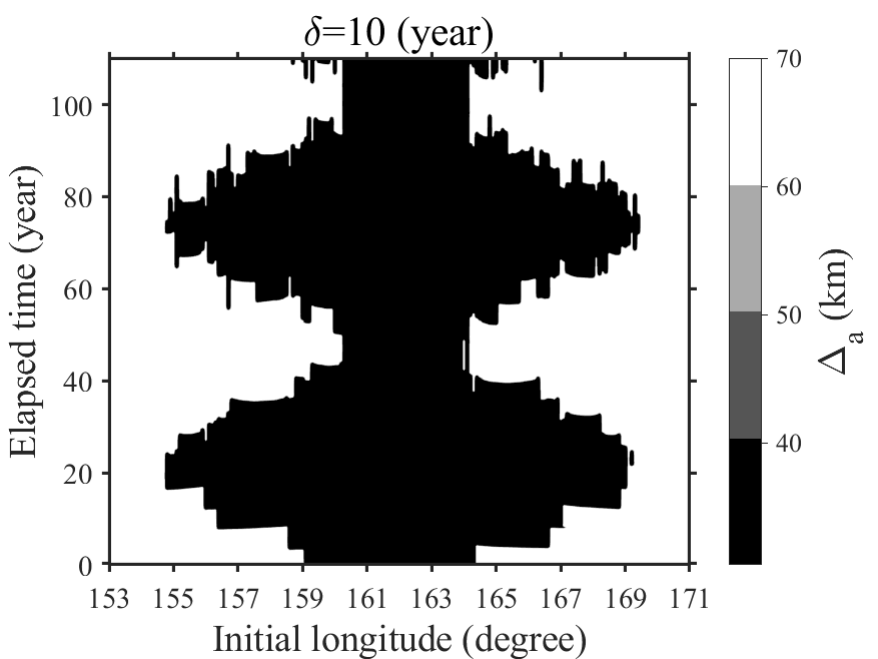}
	\caption{Semimajor axis diameter evolution with 3-year (left) and 10-year (right) window span. $t_i=$1/1/2020, N=3.}
	\label{fig_span_selection}
\end{figure}

Figure~\ref{fig_span_selection} shows the effect of using a window span too short (3 years, left pane) or too long (10 years, right pane). Because the plots are just for illustrative purposes, they use the low-fidelity model (N=3, simplified lunisolar perturbations, no Earth precession, no radiation pressure). When the window is too narrow, $\Delta_a$ fails to capture the full amplitude of the semimajor axis over a LL cycle consistently. This results in a fringe pattern that renders the plot unusable. On the other hand, if the window is too long, the areas of LL and CC are clearly delimited, but short transitions to CC are lost.

\begin{figure}[h!]
	\centering
	\includegraphics[width=0.9\textwidth]{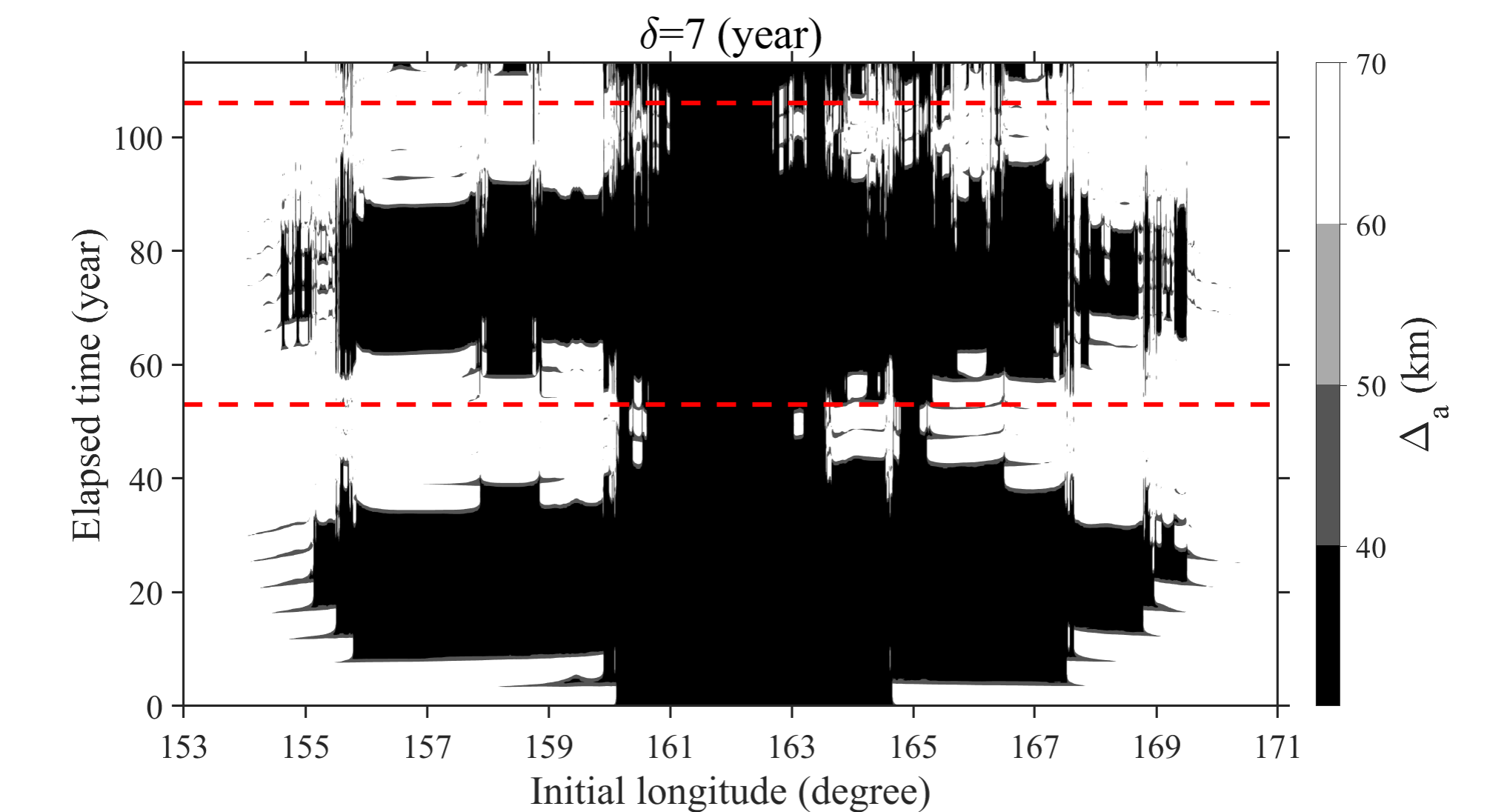}
	\caption{Semimajor axis diameter evolution with 7-year window. Red dashed lines mark 53-year cycles. $t_i=$1/1/2020, N=8.}
	\label{fig_HIFI_U2}
\end{figure}

A 7-year windows (Fig.~\ref{fig_HIFI_U2}) achieves a good compromise between robustness and fine details of the temporal evolution. In this case, the calculation uses the high fidelity model (N=8), with a resolution of 0.01$^\circ$ in initial longitude. It highlights clearly the difference between the areas of long libration (white) and continuous circulation (black). For reference, the 53-year inclination cycles have been marked with dashed red lines. It is common practice to present the gradient of the LD, instead of its actual value, to enhance the boundaries between regions of different dynamical behavior. Here, the separation is so conspicuous that this additional post-processing step becomes superfluous.

The potential barrier at the unstable point is highest when the orbit is equatorial (e.g., at the start of the propagation) and decreases as the inclination grows during the precession cycle. The minimum takes place approximately 27 years after the initial epoch (mid-cycle), and then returns to the maximum at 53 years. At the beginning of the propagation, only the trajectories starting close to $U_2$ have enough energy to move across the potential barrier and drift monotonically (see \cite{Proietti_2021} and \cite{Flores_2023} for more details). As the inclination increases, the barrier lowers and additional trajectories switch to CC. The maximum extension of the CC region occurs after 27 years, and then it begins to shrink due to the decrease in inclination.

The geometry of the boundary separating the LL and CC regions is very irregular, with the complexity growing as time increases. The transition to chaos is revealed by the sudden and irregular switches between LL and CC along a constant-time section of Fig.~\ref{fig_HIFI_U2}. They highlight how small changes in initial longitude eventually lead to vastly different motion patterns, making the trajectories diverge rapidly. For example, at the start of propagation, the range of initial longitudes undergoing CC is [160.1$^\circ$,164.7$^\circ$]. Outside that interval, all trajectories move in LL. Therefore, the initial dynamical map if highly ordered. Compare this with the situation after 90 years, when the mode of motion fluctuates irregularly for initial longitudes inside [154.3$^\circ$, 169.5$^\circ$]. Moving along a vertical line in Fig.~\ref{fig_HIFI_U2} shows that during the first transition the response is still ordered (neighboring longitudes switch mode at the same time). However, after the second transition some trajectories have become unpredictable (e.g., $\lambda_i=155.6^\circ$ at 32 years), because small variations of the initial position result in a different mode of motion.

\begin{figure}[h!]
	\centering
	\includegraphics[width=0.9\textwidth]{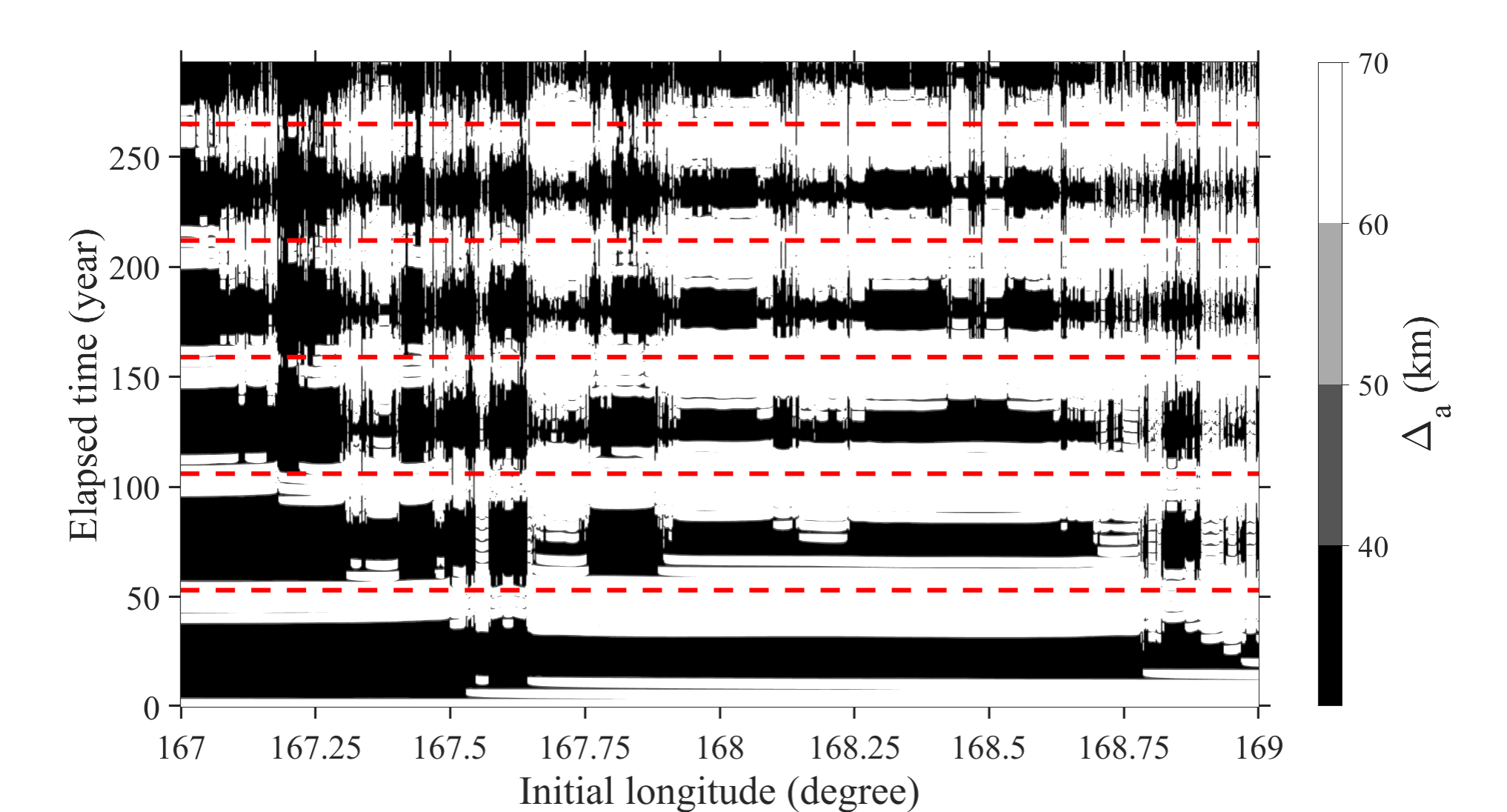}
	\caption{$\Delta_a$ evolution for $\lambda_i \in \left[ 167^\circ,169^\circ \right]$. $\delta=7$ year, $t_i=$1/1/2020, N=8.}
	\label{fig_HIFI_U2_detail}
\end{figure}

To further illustrate this chaotic behavior, Fig.~\ref{fig_HIFI_U2_detail} shows a detail between $167^\circ$E and $169^\circ$E, with 0.002$^\circ$ resolution in longitude, and the propagation extending for 300 years. It is clear that the sudden changes in dynamical state occur over very small longitude scales, and become more irregular with each precession cycle. This eventually makes the evolution unpredictable for all longitudes between $154^\circ$E and $170^\circ$E.

The horizon of unpredictability varies with the initial position, but 50 years can be taken as a rough estimate for the majority of longitudes. This is in agreement with references \cite{Breiter_2005}, \cite{Proietti_2021} and \cite{Flores_2023}. Note that there are trajectories that remain perfectly regular for very long times. Notably, those locked in CC around $\lambda_i=162^\circ$. While they transition to LL eventually (see further below for a 1000-year result), it can take several centuries. This extreme behavior was also recorded in \cite{Proietti_2021}. 

Although the evolution of individual trajectories is stochastic in practice, there is an overall periodicity of the system (the pattern repeats for each precession cycle). This is important, because the onset of unpredictability coincides with the transitions between LL and CC. For example, in \cite{Flores_2023}, which studied only one longitude close to $U_2$ ($159.3^\circ$E), it was determined that the transitions to chaos occurred near the points of null inclination of the precession cycle. Figure~\ref{fig_HIFI_U2} reveals that the switch between LL and CC shifts to higher inclinations (i.e., later in the precession cycle) as the starting position moves away from $U_2$. At the extremes of the interval where CC is possible ($\lambda_i \in$ [154$^\circ$,170$^\circ$]), transitions take place at peak inclination (i.e, midway along the precession cycle). Thus, the approximate dates when the accuracy of the predicted orbit degrades suddenly are known beforehand. They correspond to the boundary of the black region in Fig.~\ref{fig_HIFI_U2}. Therefore, orbit determination updates can be scheduled accordingly, to maintain a suitable precision. 

\begin{figure}[h!]
	\centering
	\includegraphics[width=0.9\textwidth]{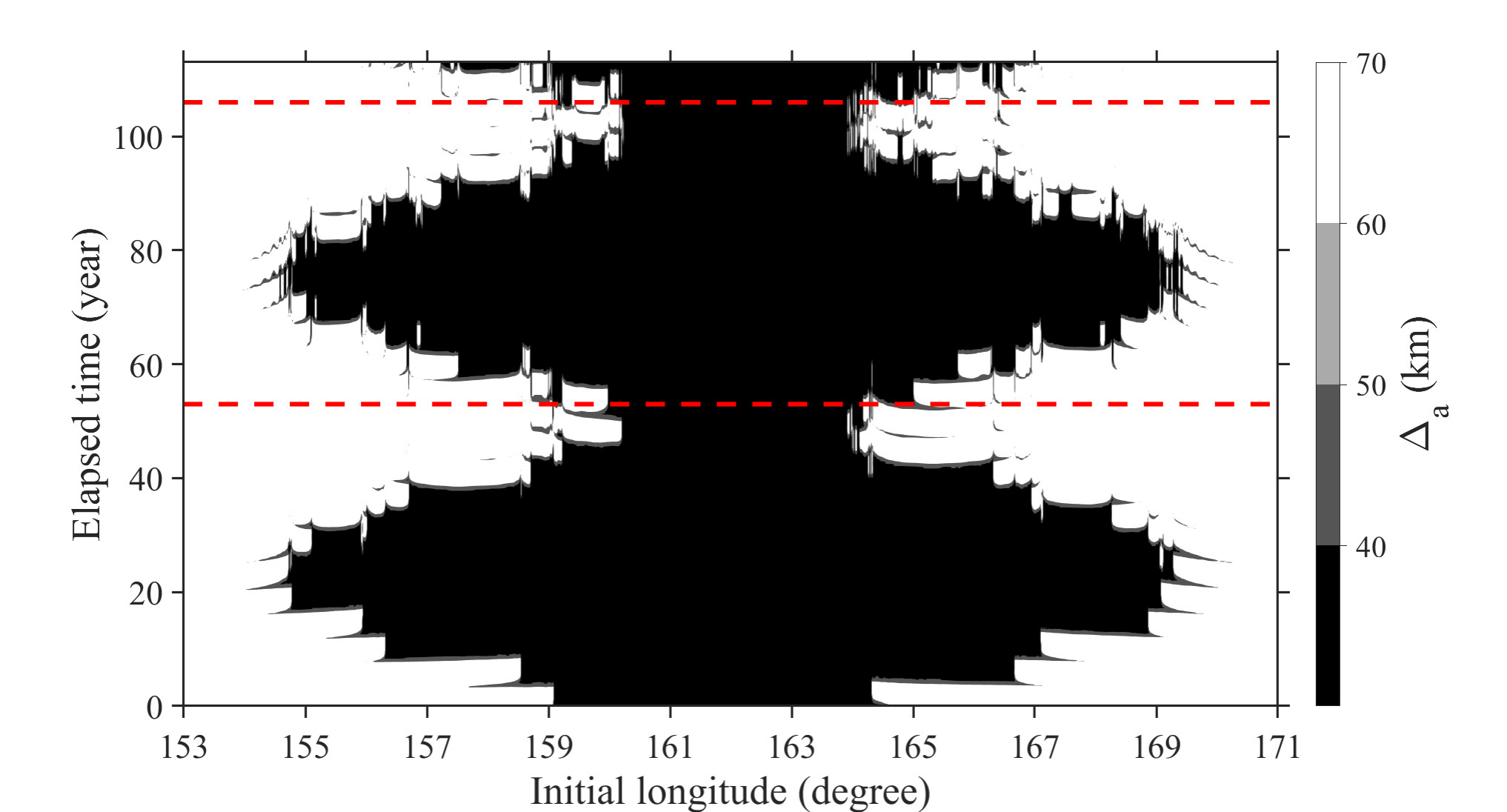}
	\caption{$\Delta_a$ evolution with low-fidelity model (N=3). $\delta=7$ year, $t_i=$1/1/2020}
	\label{fig_LOFI_U2}
\end{figure}

An important finding of \cite{Flores_2023} was that the low-fidelity model (N=3) preserves the main features of the system, also displaying chaotic response. However, that study was limited to a single initial longitude ($159.3^\circ$E) due to the high computational cost incurred by the Monte Carlo analysis. As demonstrated in Fig.~\ref*{fig_LOFI_U2}, the simplified model reproduces the qualitative behavior over the complete range of initial longitudes. It also provides a reasonable approximation for the timing of the transitions as a function of the starting position. Thus, is can help develop theoretical models of the phenomenon, as well as support the scheduling of updates to the orbit determination.

\begin{figure}[h!]
	\centering
	\includegraphics[width=0.9\textwidth]{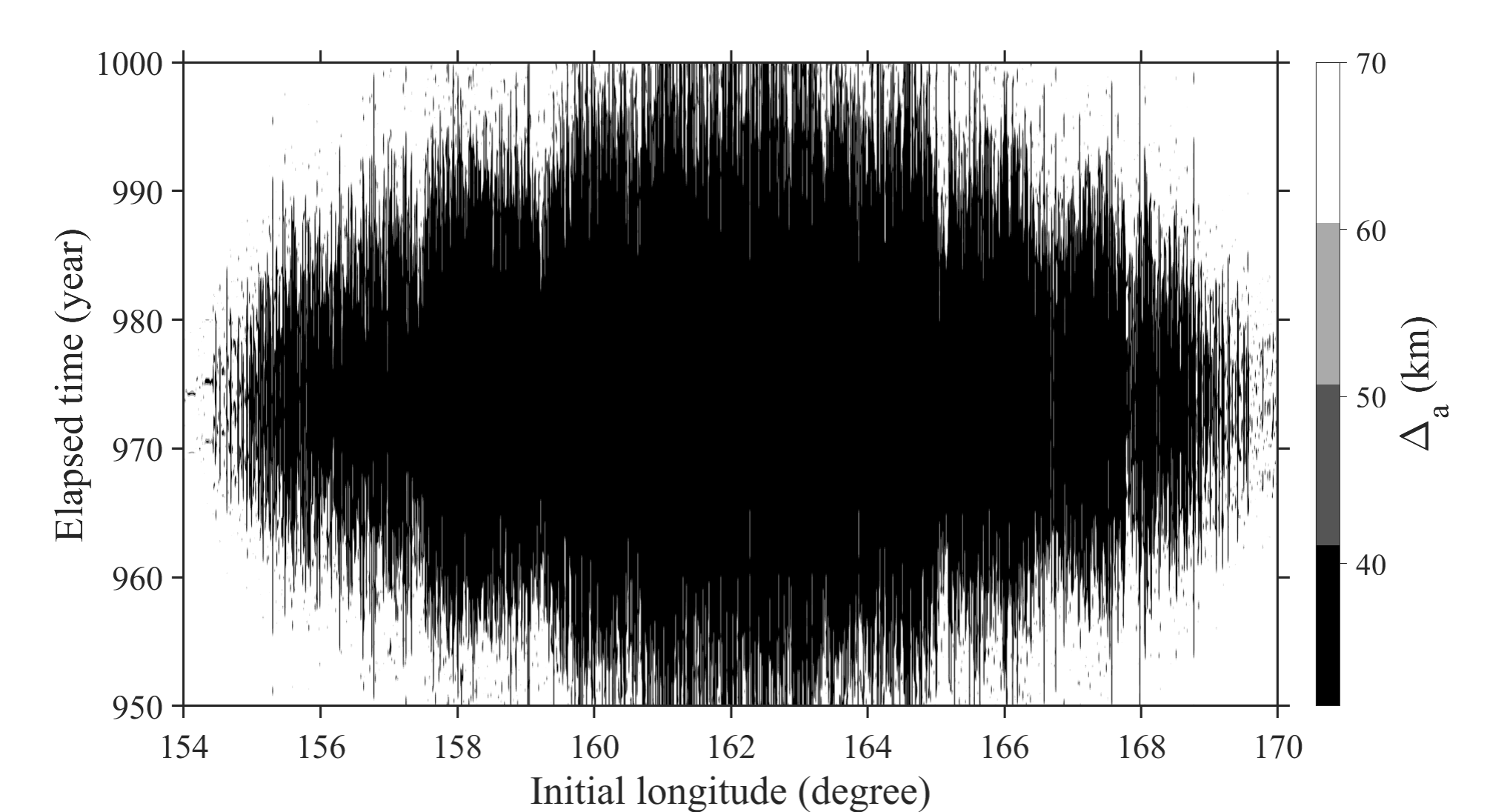}
	\caption{Long-term $\Delta_a$ evolution computed with low-fidelity model (N=3). $\delta=7$ year, $t_i=$1/1/2000.}
	\label{fig_LOFI_U2_1000y}
\end{figure}

With the low-fidelity model, the evolution towards disorder is slower (compare with Fig.~\ref*{fig_HIFI_U2}), resulting in a simpler and smoother geometry of the boundary. However, over very long time scales, the structure of the solution is also extremely complex, as shown in Fig.~\ref{fig_LOFI_U2_1000y}. It displays the last inclination cycle of a propagation that spanned 1010 years. It is clear from the fine structure of the boundary that, while the dynamics are strictly deterministic, the solution is stochastic in practice. The complete range of longitudes where transitions are possible becomes chaotic, given enough time. Note that it does not make sense to apply the high-fidelity model to this case. As indicated in Sec.~\ref{sec_num_mod}, the concise representation of the IAU 2000/2006 precession-nutation model remains accurate only for 5 centuries. Furthermore, over a scale of one millennium, the effect of $\Delta T$---the unpredictable difference between time scales based on atomic clocks and Earth's rotation \cite{Hilton_2012}---would limit the accuracy, making the high-fidelity model pointless.

\begin{figure}[h!]
	\centering
	\includegraphics[width=0.9\textwidth]{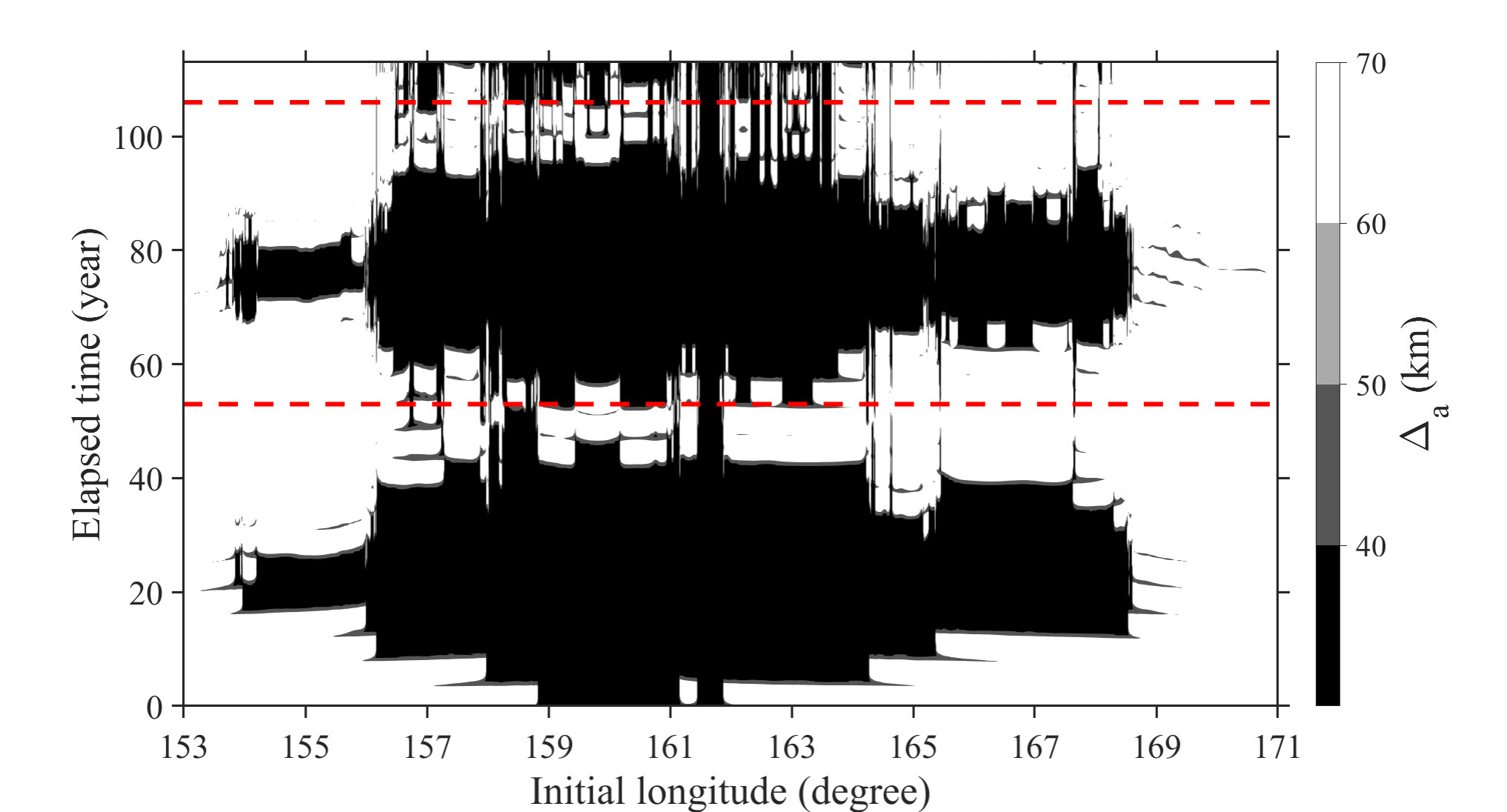}
	\caption{$\Delta_a$ evolution for initial epoch 1 June 2030. $\delta=7$ year, N=8.}
	\label{fig_HIFI_U2_jun2030}
\end{figure}

As mentioned before, the initial epoch has an effect on the evolution, because it affects the relative positions of Earth, Moon and Sun. In particular, \cite{Flores_2023} found that the inclination of the Moon's orbit relative to the equator, which varies due lunar nodal precession, affects the rate at which neighboring trajectories diverge after a transition, but does not change the main features of the system.
To check if this applies to a wider range on initial longitudes, the analysis has been repeated using 1 June 2030 at 0:00 UTC (JDN 2462653.5) as initial time. The study \cite{Proietti_2021} includes some reference data for comparison. Figure~\ref{fig_HIFI_U2_jun2030} shows the result obtained with a window of 7 years and the high-fidelity model, allowing direct comparison with Fig.~\ref{fig_HIFI_U2}. 
There are differences in the details. For example, the range of longitudes that move initially in CC is narrower and shifted west by approximately one degree. Also, at the end of the first inclination cycle ($\sim50$ years), the motion in the immediate neighborhood of $U_2$ is predominantly LL, while in Fig.~\ref{fig_HIFI_U2} it is CC.
However, the overall qualitative behavior is similar. Notably, the timing of the transitions follows the precession cycle of the orbital plane. This result confirms that most findings of \cite{Flores_2023} are applicable to the complete region where transitions between CC and LL are possible. The main difference is that the timing of the disorder episodes is now a function of the initial longitude, and only coincides with the minimum of inclination for starting positions very close to $U_2$ (which is the only situation analyzed in \cite{Flores_2023}).

\subsection{The dynamical behavior around $U_1 (15^\circ \rm{W})$}
\begin{figure}[h!]
	\centering
	\includegraphics[width=0.9\textwidth]{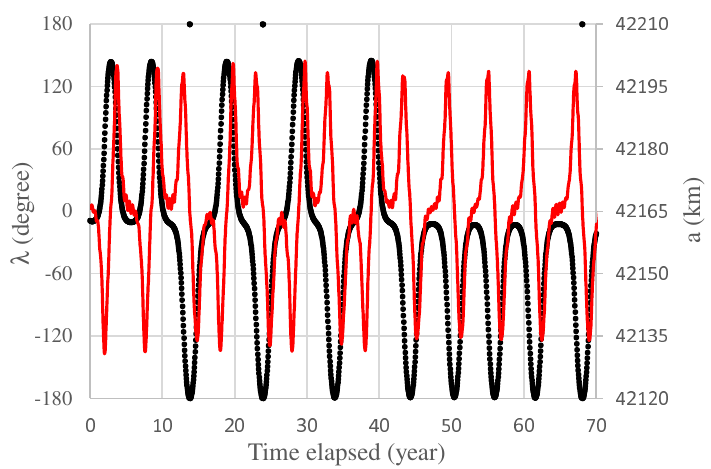}
	\caption{{Longitude (black dots) \& semimajor axis (red line) vs. time. $\lambda_i=9^\circ$W, $t_i=$1/1/2020, N=3.}}
	\label{fig_LOFI_U1_lon_t}
\end{figure}

In the case of $U_1$, the transitions occur between LL and SL modes (see Fig.~\ref{fig_LOFI_U1_lon_t} for a representative trajectory). There are two distinct short libration modes. One is confined to the vicinity of the $S_1$ stable equilibrium point (SL1 hereafter), and the other is centered on $S_2$ (SL2).

\begin{figure}[h!]
	\centering
	\includegraphics[width=0.5\textwidth]{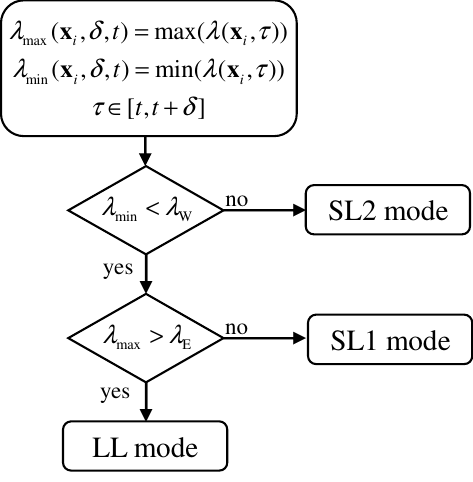}
	\caption{Logic to determine the mode of motion near $U_1$.}
	\label{fig_flowchart}
\end{figure}

The fluctuations of the semimajor axis for the LL and SL modes are similar, so $\Delta_a$ is not a reliable indicator of the dynamical state. The variable that directly correlates with the type of motion is the geographic longitude. It fluctuates approximately 160$^\circ$ during SL vs. 320$^\circ$ for LL. The diameter of the longitude ($\Delta_\lambda$) can detect the two motion types. As shown in Fig.~\ref{fig_LOFI_U1_lon_t}, the period of motion is approximately 6 years for SL and a decade for LL. Thus, with a window length between 6 and 10 years, SL corresponds to $\Delta_\lambda<180^\circ$, and diameters above $180^\circ$ signal LL. However, it is desirable to identify if the short libration mode is SL1 or SL2, to gain further insight into the dynamical behavior. This is easily accomplished with a small modification of the diameter LD. If the SRO moves across of $U_1$, the motion is marked as LL. Otherwise, if it is confined west or east of $U_1$, it is identified as SL1 or SL2, respectively. As explained before, there is no real equilibrium point once third-body perturbations are considered, but a range of longitudes where the spacecraft is in near-equilibrium. To account for this effect, a couple of threshold values $\lambda_W=-20^\circ$ and $\lambda_E=0^\circ$ are used to determine the region where the spacecraft is traveling. The logic is outlined in Fig.~\ref{fig_flowchart}. Note that a continuous representation of the longitude is required, because reducing the angle to a single turn would introduce jumps, leading to spurious identifications.

\begin{figure}[h!]
	\centering
	\includegraphics[width=0.9\textwidth]{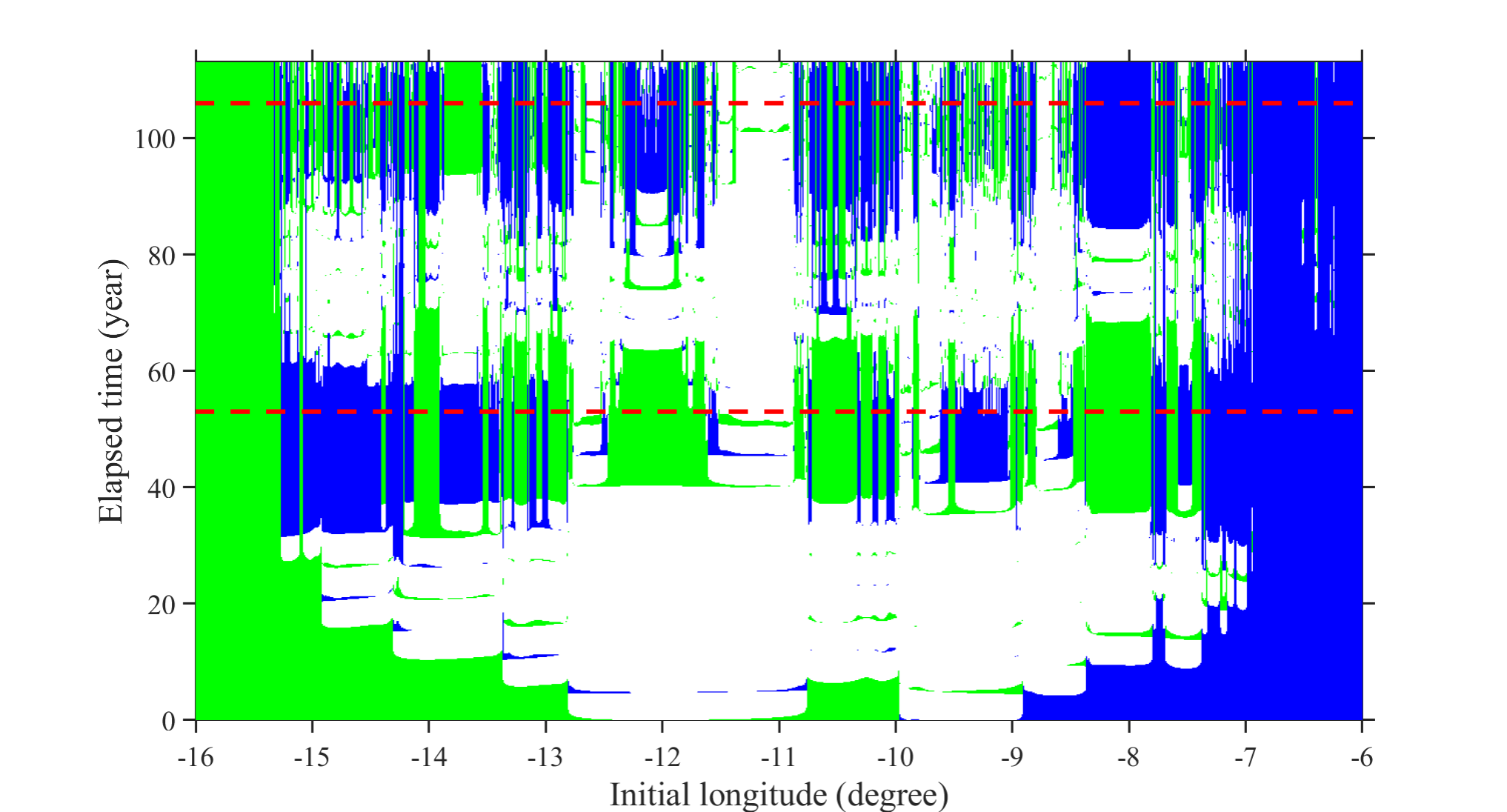}
	\caption{Transitions between LL (white), SL1 (green) and SL2 (blue). $\delta=7$ year, $t_i=$1/1/2020, N=8.}
	\label{fig_HIFI_U1_modes}
\end{figure}

Upon testing, a window size $\delta=7$ year was selected to achieve a good compromise between robustness and preservation of fine detail. The results for the first two inclination cycles, starting on 1 January 2020 (Ref.~\citealp*{Proietti_2021} contains data for this epoch), and computed with the high-fidelity model are shown in Fig.~\ref*{fig_HIFI_U1_modes}. The white areas denote LL motion, green is SL1 and blue corresponds to SL2.
Trajectories starting sufficiently far west or east of $U_1$ are confined to regular SL1 or SL2 motion, respectively. For initial longitudes between $13^\circ$W and $9^\circ$W the motion is mostly LL, but with temporary transitions to SL that become more irregular with time. As the orbital inclination increases, the barrier to move across $U_1$ is lowered, and the range of longitudes compatible with LL widens. Towards the end of the precession cycle, the LL region shrinks again. Due to the existence of three different modes, the dynamical landscape around $U_1$ is vastly more complicated than for $U_2$ (compare with Fig.~\ref{fig_HIFI_U2}). Whenever the satellite leaves the LL mode, the timing of the transition determines if it switches to SL1 or SL2. This additional flexibility leads to a much richer structure. 

\begin{figure}[h!]
	\centering
	\includegraphics[width=0.9\textwidth]{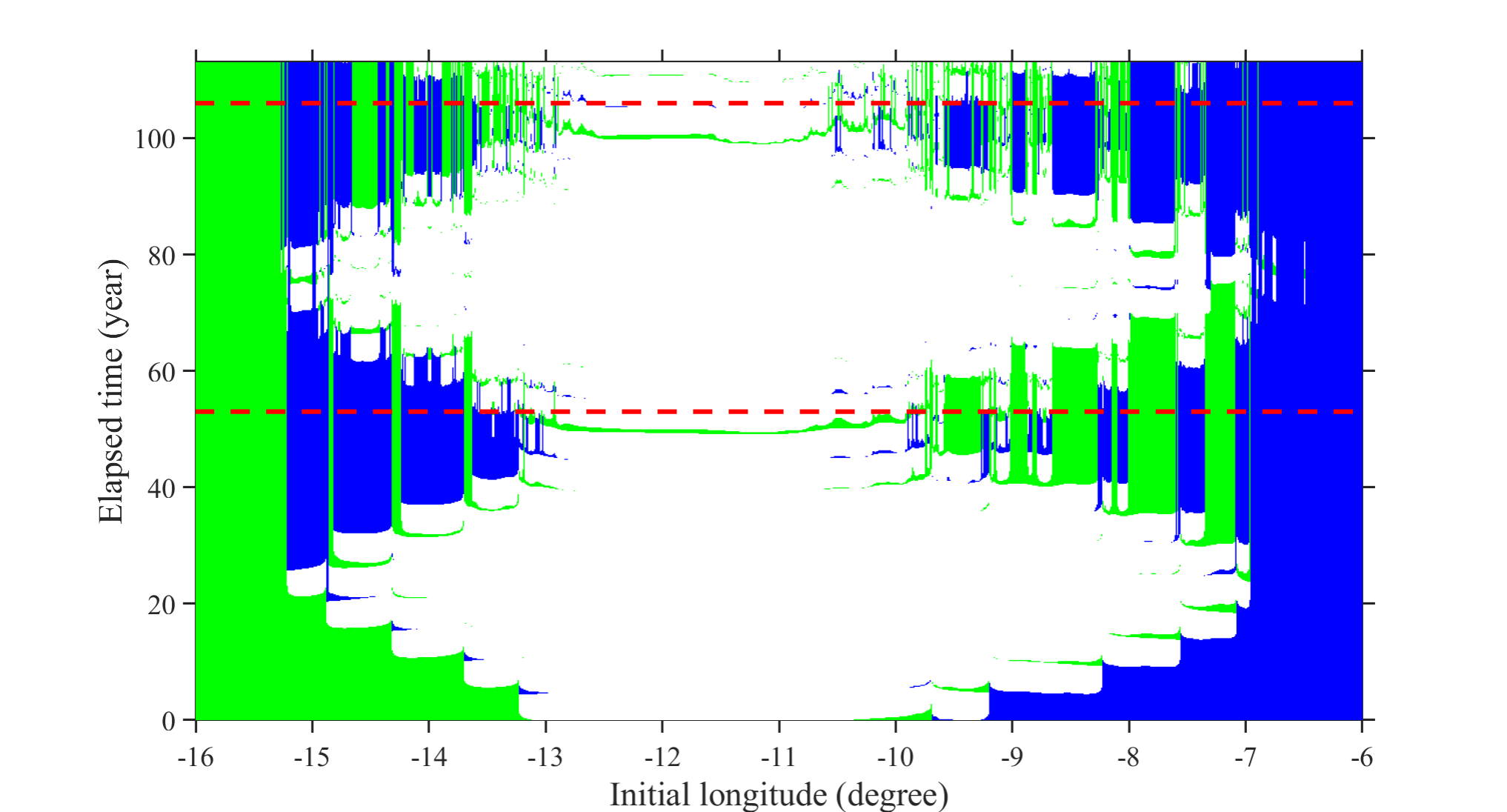}
	\caption{Transitions between LL (white), SL1 (green) and SL2 (blue) computed with low-fidelity model (N=3). $\delta=7$ year, $t_i=$1/1/2020.}
	\label{fig_LOFI_U1_modes}
\end{figure}

The low-fidelity model is able to capture the main dynamical features in the vicinity of $U_1$ (Fig.~\ref{fig_LOFI_U1_modes}). It gives a reasonable estimate of the range of initial longitudes where irregular transitions can occur. However, there is a reduction in solution complexity, as expected. For example, the shape of the LL region is much better defined, with fewer irregular transitions to SL, compared with the high-fidelity propagation.

\section{Conclusions}
\label{sec_concl}

This paper studied the dynamical behavior of uncontrolled objects in GEO, released in the neighborhood of the unstable equilibrium points of the terrestrial equator ($U_1$ and $U_2$). The long-term dynamics features transitions between continuous circulation (CC) and long libration (LL) for $U_2$, or between long libration and short libration (SL) in the case of $U_1$. It was first reported in \cite{Flores_2023} that the transitions lead to chaos, with the satellite trajectory becoming unpredictable beyond 50 years. Interestingly, the evolution towards disorder occurred only near the minimum inclination points of the 53-year orbital precession cycle, with the intervening periods displaying regular evolution. However, due to the high computational cost of the Monte Carlo analysis used, the study was limited to a single initial position in the immediate neighborhood of $U_2$. Using the Lagrangian Descriptors (LD) technique, the analysis has been extended to the complete range of initial longitudes where transitions are possible for both $U_1$ and $U_2$. A high fidelity physical model has been used, achieving the maximum accuracy possible with the EGM2008 geopotential model. It includes Earth precession and nutation, lunisolar perturbations with precomputed ephemeris and solar radiation pressure with partial eclipses. The equations of motion in Cartesian coordinates are integrated directly without any averaging.

The diameter LD (i.e., the amplitude of the variation) of the semimajor axis can discriminates between CC and LL, with the latter corresponding to a higher value. The standard approach in the literature uses a fixed time interval to evaluate the diameter, so it does not provide information about the epoch of the transitions. To overcome this limitation, a modified indicator that computes the amplitude with a moving time window is proposed. By selecting a suitable window span, the timing of the transitions can be determined for each initial longitude. The computational cost is vastly reduced compared to the Monte Carlo technique (only one propagation per initial longitude). This enables efficient mapping of the dynamical behavior over a wide range of initial positions.

For this particular problem, the LD proved superior to the largest Lyapunov exponent (LLE) indicator. Averaging the LLE over a carefully selected period, it is possible to detect changes in the mode of motion. However, it yields a noisier signal than the windowed diameter, making the results more difficult to interpret. Moreover, the LD requires half the computational effort (one propagation per initial condition, vs. two for LLE).

For $U_2$, the diameter map revealed that the timing of the transitions is always governed by the precession cycle. As the orbital inclination increases, the interval of longitudes where CC motion is possible becomes wider, with the maximum extent reached at the midpoint of the cycle (27 years after the initial epoch). Subsequently, the CC region shrinks until the point of minimum inclination, and the cycle repeats. The transitions are so sensitive to the initial conditions that it becomes impossible to predict trajectories reliably beyond 50 years. This is illustrated by the extremely complex geometry of the boundary between the CC and LL modes. Computations over one millennium evidenced that the sensitivity grows with each cycle and the long-term behavior is, in practice, stochastic.

For the $U_1$ equilibrium point, the geographical longitude is used to determine if the motion is confined to the neighborhood of one stable equilibrium point (SL1 and SL2 modes) or spans both of them (LL). Inspired by the diameter LD, a modified criterion that detects if the spacecraft crosses $U_1$ identified the three dynamical states reliably. The precise timing of the transitions from LL to SL, which is by itself unpredictable over long time scales, determines if the subsequent motion is SL1 or SL2. This additional element of complexity accelerates the evolution towards chaos. Again, the precession cycle governs the extension of the region where transitions are possible. It is wider when the inclination is near its maximum.

It has been demonstrated that the main features of the system, over the complete range of longitudes, are captured by a low-fidelity model that includes only gravity field harmonics up to degree and order 3, and lunisolar perturbations assuming circular coplanar orbits for Earth and Moon. This simplified model could be the foundation for semi-analytical studies to gain further insight into the problem.

Using customized versions of the diameter LD, the first comprehensive map of the periodic transitions to chaos in uncontrolled GEO spacecraft has been produced. It gives the approximate timing of the transitions between modes of motion, whose outcome cannot be predicted accurately, as a function of the initial longitude. This information is crucial to schedule orbit determination updates, in order to maintain reliable trajectory predictions and avert in-orbit collisions. The novel LD demonstrated their ability to map the dynamical structure of the system with limited computational cost. They are also straightforward to implement, and do not require a Hamiltonian formulation.

\section*{Funding declaration}
R. Flores, E. Fantino and H. Susanto acknowledge Khalifa University of Science and Technology's internal grant CIRA-2021-65/8474000413. E. Fantino received funding from project 8434000368 of the UAE Space Agency (6U Cubesat Mission) and project 8434000634 of the Polar Research Center-Presidential Court of the UAE. R. Flores and E. Fantino acknowledge grant ELLIPSE/8434000533 (Abu Dhabi's Technology Innovation Institute). E. Fantino has been partially funded by projects PID2021-123968NB-I00 and PID2020-112576GB-C21 (Spanish Ministry of Science and Innovation).

\bibliography{References}

\renewcommand{\theequation}{A.\arabic{equation}}
\setcounter{equation}{0}

\section*{APPENDIX: Governing equations}
This is a very concise summary of the physical model. For more details, use the references provided.

The state vector ${\bf x} = [{\bf r}, {\bf v}]^T$, where $\bf r$ denotes position and $\bf v$ velocity, evolves in time according to:
\begin{equation} \label{eq_motion2}
	{\bf \dot x} = {\bf f} ({\bf x}, t) \qquad t \in [t_i, t_f],
\end{equation}
\begin{equation} \label{eq_inicon2}
	{\bf x} (t_i) = {\bf x}_i.
\end{equation}
The right-hand side of \ref{eq_motion2} is ${\bf f} = [{\bf v}, {\bf a}]^T$, with $\bf a$ denoting the acceleration in the non-rotating Geocentric Celestial Reference System (GCRS). The acceleration contains contributions due to Earth (subscript E), Moon (M) and Sun (M) gravity, as well as solar radiation pressure (SRP):
\begin{equation} \label{eq_acc_comp}
	{\bf a} = {\bf a}_E + {\bf a}_M + {\bf a}_S + {\bf a}_{SRP}
\end{equation}

\subsection*{Harmonic synthesis of the gravitational field}
Earth's gravitational acceleration is the gradient of the geopotential ($V$). In spherical coordinates $\{r,\theta ,\lambda\}$ (radius, colatitude, longitude), such that $\{x,y,z\} = \{r \sin \theta \cos \lambda,r \sin \theta \sin \lambda,r \cos \theta\}$,
\begin{equation} \label{eq_acc_pot}
	{\bf a}_E = \frac{\partial V}{\partial r} {\bf{e}}_r + \frac{\partial V}{\partial \theta} \frac{{\bf e}_\theta}{r} + \frac{\partial V}{\partial \lambda} \frac{{\bf e}_\lambda}{r\sin \theta} .
\end{equation}
Expression~\ref{eq_acc_pot} uses the common convention in geodesy that the acceleration is the positive gradient of $V$. The potential is computed as a sum of spherical harmonics \cite{Holmes_2002}
\begin{equation} \label{eq_acc_sph}
	V(r,\theta ,\lambda) = \frac{\mu_E}{r} \left( 1 + \sum \limits_{n = 2}^N \left( \frac{a_E}{r} \right)^n \Omega_n \right) ,
\end{equation}
where $N$ is the expansion degree, $\mu_E$ and $a_E$ denote Earth's gravitational parameter and equatorial radius, and
\begin{equation} \label{eq_omega}
	\Omega _n = \sum\limits_{m = 0}^n \bar P_{nm} (\theta) \left( \bar C_{nm}^1 \cos m\lambda  + \bar C_{nm}^2 \sin m\lambda \right) . 
\end{equation}
In Eq.~\ref{eq_acc_sph}, the sum starts at 2 because it assumes that the origin of coordinates coincides with Earth's barycenter. In this case, the terms of degree 1 vanish. $\bar P_{nm} (\theta)$ in Eq.~\ref{eq_omega} denotes the fully-normalized associated Legendre function of the first kind of degree n and order m, and $\{C_{nm}^1,C_{nm}^2\}$ are the corresponding Stokes coefficients of the geopotential model (EGM2008 \cite{Pavlis_2012}\cite{IERS_2010a}). The gradient of the potential is evaluated with the modified forward row recursion scheme of Holmes et al. \cite{Holmes_2002}. 

\subsection*{Sun and Moon gravity}
The gravitational perturbation from a third body is given by its tidal force. That is, the difference between the gravitational pulls it exerts on the orbiter and on the origin of coordinates (Earth’s barycenter). For example, in the case of the Moon:
\begin{equation} \label{eq_gra_moon}
	{\bf a}_M = \mu_M \left( \frac{{\bf x}_M - {\bf x}}{\left\| {\bf x}_M - {\bf x} \right\| ^3}  - \frac{{\bf x}_M}{\left\| {\bf x}_M \right\| ^3} \right) .
\end{equation}
Unfortunately, Eq.~\ref{eq_gra_moon} is ill-conditioned because it contains the difference of two very similar vectors ($\left\| {\bf x} \right\| << \left\| {\bf x}_M \right\|$, this is even more problematic for the Sun). Evaluating the difference with finite precision arithmetic could cause catastrophic cancellation, resulting in large rounding errors. This issue is mitigated reworking the expression of the tidal acceleration into \cite{Battin_1999}
\begin{equation}
	{\bf a}_M = \frac{-\mu_M}{\left\| {\bf x}_M \right\|^3 (q + 1)^{3/2}} \left( {\bf x} + f {\bf x}_M \right) ,
\end{equation}
where
\begin{equation}
	q = \frac{\left\| {\bf x} \right\|^2 - 2 {\bf x} \cdot {\bf x}_M}{\left\| {\bf x}_M \right\|^3} \quad , \quad  f = q \frac{3 + 3q + q^2}{1 + (q + 1)^{3/2}} .
\end{equation}
The geocentric positions of the Moon and the Sun are reconstructed from ephemeris tables generated by the NASA-JPL Horizons System \cite{JPL_hor}. The tabular data is interpolated with cubic splines.

\subsection*{Solar radiation pressure}
The orbiting object (denoted with subscript o) is treated as a sphere ("canonball" model \cite{Kubo_1999}) with 100\% specular reflectivity. The acceleration due to radiation pressure is 
\begin{equation}
{\bf a}_{SRP} =  - \phi \frac{L_S}{4\pi c} \frac{A_\text o}{m_\text o} \frac{{\bf x}_{{\text o}S}}{\left\| {\bf x}_{{\text o}S} \right\|^3} ,
\end{equation}
where $L_S$ is the bolometric luminosity of the Sun ($3.83 \times 10^{26}$W), $c$ is the speed of light, $A_\text o / m_\text o$ is the area-to-mass ratio of the object and ${\bf x}_{{\text o}S} = {\bf x}_{S} - {\bf x}$. The factor $\phi$ accounts for eclipses, it is the fraction of the solar disk that is visible. To calculate $\phi$, start by determining the apparent angular radii of Earth and Sun as seen from the orbiter:
\begin{equation}
	\alpha_E = \arcsin \left( \frac{a_E}{\left\| {\bf x}_{{\text o} E} \right\|} \right) \quad , \quad \alpha_S = \arcsin \left( \frac{a_S}{\left\| {\bf x}_{{\text o} S} \right\|} \right) ,
\end{equation}
with $a_S = 6.97 \times 10^{5}$km denoting the solar radius and ${\bf x}_{{\text o}E} = {\bf x}_{E} - {\bf x}$. The angular separation between Earth and Sun is
\begin{equation}
	\beta  = \arccos \left( \frac{{\bf x}_{{\text o}E} \cdot {\bf x}_{{\text o}S}} {\left\| {\bf x}_{{\text o}E} \right\| \left\| {\bf x}_{{\text o}S} \right\|} \right) ,
\end{equation}
which gives rise to three distinct cases
\begin{eqnarray}
	\beta > (\alpha_E + \alpha_S) \to  \phi = 1 \quad \text{(no eclipse)} \nonumber \\
	(\alpha_E - \alpha_S) < \beta < (\alpha_E + \alpha_S) \to  \phi \in ]0,1[ \quad \text{(partial eclipse)}  \nonumber \\
	\beta < (\alpha_E - \alpha_S) \to  \phi = 0 \quad \text{(total eclipse)}  \nonumber
\end{eqnarray}
For the partial eclipse case (object in penunbra), the exact expression of $\phi$ (the relative area of a lune) is cumbersome and costly to evaluate. For an object orbiting Earth, it is safe to assume $\alpha_E >> \alpha_S$, which leads to a simpler form \cite{Flores_2021}
\begin{equation}
	\phi  \approx 1 + \frac{\varphi \sqrt {1 - \varphi ^2}  - \arccos \varphi}{\pi } \quad , \quad \varphi  = (\beta  - \alpha _E) / \alpha _S
\end{equation}

\subsection*{Orientation of the terrestrial reference frame}
While the acceleration is integrated in GCRS, the geopotential is expressed in the Earth-fixed Terrestrial Reference System (TRS). To compute the acceleration due to Earth's gravity, the transformation between the two systems is required. It is given by \cite{Hilton_2012}
\begin{equation} \label{eq_ref_trans}
	{\bf x}_\text{TRS} = \bf{W}(t) \: {\bf R}_z(\theta) \: {\bf C}(t) \: {\bf x}_\text{GCRS}.
\end{equation}
In Eq.~\ref{eq_ref_trans} ${\bf C}(t)$ denotes the combined bias-precession-nutation matrix, which accounts for changes of orientation of the axis of rotation of the Earth. It has been computed with a concise representation of the IAU 2000/2006 model \cite{Hilton_2006}, based on \cite{Capitaine_2008}. It is accurate to 1 arcsec up to 500 years into the future. ${\bf C}(t)$ transforms between GCRS and the Celestial Intermediate Reference System (CIRS), a quasi-inertial reference with the z axis aligned with Earth's poles (i.e., it does not include the diurnal rotation). Next, ${\bf R}_z(\theta)$ is a rotation about z (i.e.,Earth's axis) through an angle $\theta$ (called the Earth Rotation Angle, ERA). ERA is a linear function of the UT1 time standard, it increases at a rate of 1.00273781191135448 revolutions each 86400 seconds \cite{IERS_2010b}. The relationship between UT1, based on Earth's rotation, and the atomic standard TT (Terrestrial Time), which measures physical time, is
\begin{equation} \label{eq_ut1}
	UT1 = TT - \Delta T ,
\end{equation}
where $\Delta T$ accounts for irregularities in Earth's angular velocity. It must be determined experimentally by observing distant celestial objects, because it cannot be predicted reliably. Therefore $\Delta T$ has been frozen at its value at the beginning of 2025. The latest data con be found at the IERS website \cite{IERS_web}.
Finally, matrix $\bf{W}(t)$ accounts for polar motion, which is the displacement of Earth's axis with respect to the crust. It is also unpredictable, so it is assumed that $\bf{W}(t) = {\bf I}$. This introduces an uncertainty of 0.3 arcsec in the calculations \cite{Hilton_2012}.

\end{document}